\documentclass[11pt,letterpaper]{amsart}

\usepackage[foot]{amsaddr}
\usepackage{amsmath}
\usepackage{amsthm}
\usepackage{amssymb}
\usepackage{url}
\usepackage{hyperref}
\usepackage{mathtools}
\usepackage[left=1in,right=1in,top=1in,bottom=1in]{geometry}
\pdfoutput=1

\newtheorem{Thm}{Theorem}
\newtheorem{Lem}[Thm]{Lemma}

\theoremstyle{remark}

\theoremstyle{definition}

\newenvironment{Proof}{\begin{proof}}{\end{proof}}

\newcommand{\bra}{\ensuremath{\langle}}
\newcommand{\ket}{\ensuremath{\rangle}}
\newcommand{\per}{\ensuremath{\operatorname{per}}}
\newcommand{\dlog}{\ensuremath{\operatorname{dlog}}}
\newcommand{\iv}[1]{\ensuremath{[\![{#1}]\!]}}

\newcommand{\bbiv}[1]{\ensuremath{\biggl[\!\!\biggl[{#1}\biggr]\!\!\biggr]}}

\begin{document}


\title[]{Counting Short Vector Pairs by Inner Product\\
         and Relations to the Permanent}        

\author{Andreas Bj\"orklund}
\address{This work was carried out while AB was employed as a researcher at Lund University, Department of Computer Science,
and the major part of the writeup was carried out while AB was employed as a researcher at Ericsson Research.}
\email{andreas.bjorklund@yahoo.se}

\author{Petteri Kaski}
\address{Aalto University, Department of Computer Science, Finland}
\email{petteri.kaski@aalto.fi}

\begin{abstract}
Given as input two $n$-element sets $\mathcal A,\mathcal B\subseteq\{0,1\}^d$ 
with $d=c\log n\leq(\log n)^2/(\log\log n)^4$ and a target 
$t\in \{0,1,\ldots,d\}$, we show how to count the number of pairs 
$(x,y)\in \mathcal A\times \mathcal B$ with integer inner product 
$\langle x,y \rangle=t$ deterministically, in 
$n^2/2^{\Omega\bigl(\!\sqrt{\log n\log \log n/(c\log^2 c)}\bigr)}$ 
time.
This demonstrates that one can solve this problem in deterministic 
subquadratic time almost up to $\log^2 n$ dimensions, nearly matching 
the dimension bound of a subquadratic randomized detection algorithm
of Alman and Williams [FOCS~2015]. We also show how to modify
their randomized algorithm to count the pairs w.h.p., 
to obtain a fast randomized algorithm.

Our deterministic algorithm builds on a novel technique of reconstructing 
a function from sum-aggregates by prime residues, which can be seen as 
an {\em additive} analog of the Chinese Remainder Theorem.

As our second contribution, we relate the fine-grained complexity of the
task of counting of vector pairs by inner product to the task of computing 
a zero-one matrix permanent over the integers.
\end{abstract}

\maketitle


\section{Introduction}

\subsection{The Inner Product and the Size of Preimages}
The inner product map $\bra x,y\ket=\sum_{i=1}^d x_iy_i$ of two
$d$-dimensional vectors $x=(x_1,x_2,\ldots,x_d)$ and $y=(y_1,y_2,\ldots,y_d)$
is one of the cornerstones of linear algebra and its applications.
For example, when $x$ and $y$ are vectors of observations normalized to
zero mean and unit standard deviation, then $\bra x,y\ket$ is
the Pearson correlation between $x$ and $y$. As such, it is a fundamentally
important computational and data-analytical task to efficiently gain 
information about the {\em preimages} of the inner product 
map; for example, to highlight pairs of similar or dissimilar observables
between two families of $n$ observables.

Accordingly, the protagonist of this paper is the following counting problem 
(\#\textsc{InnerProduct}):
\begin{quote}
Given as input a target $t\in\{0,1,\ldots,d\}$ and two $n$-element sets 
$\mathcal A\subseteq \{0,1\}^d$ and $\mathcal B\subseteq \{0,1\}^d$, 
count the number of vector pairs $(x,y)\in \mathcal A \times \mathcal B$ 
with integer inner product $\langle x,y \rangle=t$.
\end{quote}
From a complexity-theoretic standpoint, this  problem generalizes many 
conjectured-hard problems in the study of fine-grained complexity---such as 
the $t=0$ special case, the {\em orthogonal vector counting} (\#\textsc{OV}) 
problem---as well as generalizing fundamental application settings, such as 
similarity search in Hamming spaces. While it is immediate that subquadratic 
scalability in $n$ is obtainable when $d=o(\log n)$, our interest in this paper 
is to obtain an improved understanding of the fine-grained complexity landscape 
for {\em moderately short} vectors, specifically for $d$ at most 
poly-logarithmic in $n$. 

\subsection{Subquadratic Scaling for Moderately Short Vectors}

Our main positive result establishes deterministic subquadratic scalability 
for \#\textsc{InnerProduct} up to $d$ growing essentially as the square of 
the logarithm of $n$: 
\begin{Thm}[Main; Subquadratic Scaling for \#\textsc{InnerProduct}]
\label{thm:main}
There exists a deterministic algorithm that, given as input 
a target $t\in\{0,1,\ldots,c\log n\}$ and 
two $n$-element sets $\mathcal{A},\mathcal{B}\subseteq\{0,1\}^{c\log n}$ 
with $4\leq c\leq\frac{\log n}{(\log\log n)^4}$, 
outputs the number of pairs $(x,y)\in\mathcal{A}\times\mathcal{B}$ 
with $\bra x,y\ket=t$ in time
\begin{equation}
\label{eq:main-time}
n^{2}/2^{\Omega\bigl(\sqrt{\frac{\log n\log\log n}{c\log^2 c}}\bigr)}\,.
\end{equation}
\end{Thm}

The algorithm in Theorem~\ref{thm:main} is based on a novel technique of 
reconstructing a function from its sum-aggregates by prime residue, which 
can be seen as an {\em additive} analog of the Chinese Remainder Theorem 
and may be of independent interest (cf.~Sect.~\ref{sect:reconstruction}).

We also show how a randomized algorithm for the decision problem 
of checking for a pair of vectors whose Hamming distance is less than a target
by Alman and Williams~\cite{AlmanW2015}, can with a small modification be turned into an algorithm for \#\textsc{InnerProduct}.

\begin{Thm}[Randomized Subquadratic Scaling for \#\textsc{InnerProduct}]
\label{thm:random}
There exists a randomized algorithm that w.h.p., given as input 
a target $t\in\{0,1,\ldots,c\log n\}$ and 
two $n$-element sets $\mathcal{A},\mathcal{B}\subseteq\{0,1\}^{c\log n}$ 
with $4\leq c\leq\frac{\log n}{(\log\log n)^3}$, 
outputs the number of pairs $(x,y)\in\mathcal{A}\times\mathcal{B}$ 
with $\bra x,y\ket=t$ in time
\begin{equation}
\label{eq:random-time}
n^{2}/2^{\Omega\bigl(\frac{\log n}{c\log^2 c}\bigr)}\,.
\end{equation}
\end{Thm}

While the randomized algorithm in Theorem~\ref{thm:random} is 
faster than the deterministic one in Theorem~\ref{thm:main}, we
stress that as far as we know no deterministic algorithm in subquadratic 
time was previously known for \#\textsc{InnerProduct}, even for 
$O(\log n)$ dimensions. In particular, derandomizing Theorem~\ref{thm:random} 
while retaining subquadratic time seems challenging, even though some 
progress on the amount of randomness needed in the algorithm has been made, 
cf.~Theorem~1.1~in~\cite{AlmanCW2016}. 

Our further objective is to better understand the fine-grained complexity 
of \#\textsc{InnerProduct} in relation to that of \#\text{OV} and other counting
problems.
For $d=O(\log n)$, it is known that these problems are truly-subquadratically 
related; indeed, Chen and Williams~\cite{ChenW19} give a parsimonious 
reduction for the detection variants of these two problems. 
That is, if \#\textsc{OV} can be solved in $n^{2-\omega(1)}$ time, 
then so can \#\textsc{InnerProduct}. However, while there is a 
subquadratic time algorithm for \#\textsc{OV} whose running time scales as good as $n^{2-\Omega(1/\log c)}$~\cite{ChanW2016}, 
the reduction of Chen and Williams~\cite{ChenW19} does not immediately give 
a non-trivial algorithm for \#\textsc{InnerProduct}. Indeed, the fastest known
algorithm for the decision version \textsc{InnerProduct} utilize probabilistic 
polynomials for symmetric Boolean functions with optimal dependence on 
the degree and error~\cite{AlmanW2015}, and does not go via fast \textsc{OV} algorithms and 
the reduction above. In Theorem~\ref{thm:random}, we show how a simple modification to the algorithm in Alman and Williams~\cite{AlmanW2015}
can turn their algorithm into a counting one. We note that while Alman, Chan, and Williams~\cite{AlmanCW2016} later presented a deterministic algorithm based on Chebyshev polynomials over the reals for minimum/maximum Hamming weight pair, with the same running time as the randomized one in~\cite{AlmanW2015}, that deterministic algorithm, or the even faster randomized one they presented, can not be turned into one for \#\textsc{InnerProduct}
by our suggested modification alone.

\subsection{Lower Bounds via the Permanent}

The running times~\eqref{eq:main-time} and ~\eqref{eq:random-time} would, 
at least at first, appear to leave room for improvement. Indeed, the 
running time~\eqref{eq:random-time} is considerably worse than the 
running time $n^{2-\Omega(1/\log c)}$ obtained by 
Chan and Williams~\cite{ChanW2016} for \#\textsc{OV}. We proceed to show that 
this intuition might be misleading, since such scalability would imply the 
existence of considerably faster algorithms for a canonical hard problem 
in exponential-time complexity. 
Accordingly, 
to gain insight into the complexity of \#\textsc{InnerProduct} and \#\textsc{OV}
when $d=\omega(\log n)$, we introduce our second protagonist 
($R$-\textsc{Permanent}):
\begin{quote}
Given as input an $n\times n$ matrix $M$ with entries $m_{ij}$ in 
a ring $R$ for $i,j\in[n]$, compute the {\em permanent}
\[
\per M=\sum_{\sigma\in S_n} \prod_{i\in [n]} m_{i,\sigma(i)}\,,
\]
where $S_n$ is the group of all permutations of $[n]=\{1,2,\ldots,n\}$.
\end{quote}
Ryser's algorithm from 1963 computes the permanent with $O(n2^n)$ arithmetic 
operations in $R$~\cite{Ryser1963}. It is a major open problem whether 
this can be improved to $O(c^n)$ for some constant $c<2$. Even improving 
the running time to less than $2^n$ operations has been noted as 
a challenge by Knuth in the 
{\em Art of Computer Programming} \cite[Exercise 4.6.4.11]{Knuth1998}.
Valiant in 1979 famously proved that the permanent is \#P-complete even 
when restricted to $m_{ij}\in\{0,1\}$ and evaluated over the ring of 
integers~\cite{Valiant1979}; this version of the problem can be interpreted 
as counting the perfect matchings in a balanced bipartite graph having 
the matrix as its biadjacency matrix.
For zero-one inputs over the integers, \emph{somewhat} faster algorithms are 
known (cf.~Sect.~\ref{sect:related}); to the best of our knowledge,
the current champion for zero-one matrices computes the permanent 
in $2^{n-\Omega\bigl(\sqrt{n/\log \log n}\bigr)}$ time~\cite{BjorklundKW2019}.

As our second contribution, we relate the fine-grained scalability 
of solving \#\textsc{InnerProduct} and \#\textsc{OV} to the task of computing the 
permanent of a zero-one matrix over the integers. In particular, our first 
result shows that if we could solve \#\textsc{InnerProduct} as fast as the
fastest currently known algorithms for \#\textsc{OV}~\cite{ChanW2016}, 
then we would immediately obtain a much faster algorithm for the permanent: 

\begin{Thm}[Lower Bound for \#\textsc{InnerProduct} via Integer Permanent]
\label{thm:perm-to-exactip}
If there exists an algorithm for solving \#\textsc{InnerProduct} for 
$N$ vectors from $\{0,1\}^{c\log N}$ in time $N^{2-\Omega(1/\log c)}$, 
then there exists an algorithm solving the permanent of an $n\times n$ 
zero-one matrix over the integers in time $2^{n-\Omega(n/\log n)}$.
\end{Thm}

Thus, despite the true-subquadratic equivalence for $d=O(\log n)$~\cite{ChenW19}, 
it would appear that \#\textsc{InnerProduct} and \#\textsc{OV} have 
different complexity characteristics when $d=\omega(\log n)$.

Our next result shows that a modest improvement in fine-grained 
scalability of \#\textsc{OV} would likewise imply much faster algorithms 
for the permanent. 

\begin{Thm}[Lower Bound for \#\textsc{OV} via Integer Permanent]
\label{thm:perm-to-ov}
If there exists an algorithm for solving \#\textsc{OV} for 
$N$ vectors from $\{0,1\}^{c\log N}$ in time 
$N^{2-\Omega(1/\log^{1-\epsilon} c)}$ for some $\epsilon>0$, then there
exists an algorithm solving the permanent of an $n\times n$ 
zero-one matrix over the integers in time 
$2^{n-\Omega(n/\log^{2/\epsilon-2} n)}$.
\end{Thm}

We note that such fast algorithms for \#\textsc{OV} would already disprove the 
so-called Super Strong ETH, that $k$-\textsc{CNFSAT} on $n$ variables 
has a $2^{n-n/o(k)}$ time algorithm, by the reduction to \textsc{OV} 
by Williams~\cite{Williams2005} after sparsification \cite{ImpagliazzoPZ2001}. The present result merely adds to the list of consequences of faster algorithms for \#\textsc{OV}.

\subsection{Methodology and Organization of the Paper}

The key methodological contribution underlying our main algorithmic result 
(Theorem~\ref{thm:main}) is a novel additive analog of the Chinese Remainder
Theorem (Lemma~\ref{lem:reconstruction} developed independently of the
application in Sect.~\ref{sect:reconstruction}), which enables us to recover 
the number of pairs 
$(x,y)\in\mathcal{A}\times\mathcal{B}$ with $\bra x,y\ket=t$ from counts of
pairs $(x,y)$ satisfying $\bra x,y\ket\equiv r\pmod p$ for multiple  
small primes $p$ and residues $r\in\{0,1,\ldots,p-1\}$. In particular, the crux
of the algorithmic speedup lies in the observation 
that to recover the count associated with a target $0\leq t\leq d$, primes 
up to roughly $\sqrt{d}$ suffice by Lemma~\ref{lem:reconstruction}.
To obtain the counts of pairs in each residue
class $r$ modulo $p$, we employ the polynomial method with modulus-amplifying 
polynomials of Beigel and Tarui~\cite{BeigelT1994} to accommodate the counts 
under a prime-power modulus, with fast rectangular matrix multiplication of 
Coppersmith~\cite{Coppersmith1982} as the key subroutine implementing 
the count; this latter part of the algorithm design developed 
in Sect.~\ref{sect:main-proof} follows well-known techniques in fine-grained 
algorithm design (e.g.~\cite{AlmanCW2016}). Similarly, the
randomized algorithm design in Theorem~\ref{thm:random} follows by a minor 
adaptation of the probabilistic-polynomial techniques of Alman and 
Williams~\cite{AlmanW2015} to a counting context; a proof is relegated 
to Sect.~\ref{sect:proof-of-randomized-algorithm}.

Our two lower-bound reductions, Theorem~\ref{thm:perm-to-exactip} 
and Theorem~\ref{thm:perm-to-ov}, rely on reducing an $m\times m$
integer permanent first via the Chinese Remainder Theorem into
permanents modulo multiple primes $p$ with $p\leq m\ln m$, and then using 
algebraic splitting via Ryser's formula~\cite{Ryser1963} to obtain 
short-vector instances of \#\textsc{InnerProduct} and \#OV, respectively. 
For \#\textsc{InnerProduct} and Theorem~\ref{thm:perm-to-exactip},
the split employs a novel discrete-logarithm version of Ryser's formula 
modulo $p$ to arrive at two collections of vectors whose counts of pairs with
specific inner products enable recovery of the permanent modulo $p$;
the proof is presented in Sect.~\ref{sect:perm}.
For \#OV and Theorem~\ref{thm:perm-to-ov}, the split analogously 
employs Ryser's formula modulo $p$ but with a more intricate vector-coding
of group residues modulo $p$ to obtain the desired correspondence with counts 
of pairs of orthogonal vectors; we relegate the proof to 
Sect.~\ref{sect:perm-cont}.

\subsection{Related Work and Further Applications}
\label{sect:related}
{\em Exact and approximate inner products.} 
Abboud, Williams, and Yu~\cite{AbboudWY2015} used the polynomial method 
to construct a randomized subquadratic time algorithm for 
\textsc{OV}. Chan and Williams~\cite{ChanW2016} derandomized the algorithm 
and showed that it could also solve the counting problem \#\textsc{OV}. 
The first result that addressed an inner product different from zero, was 
the randomized algorithm for minimum
Hamming weight pair by Alman and Williams~\cite{AlmanW2015}. Subsequently,
Alman, Chan, and Williams~\cite{AlmanCW2016} found an even faster randomized  
as well as a deterministic subquadratic algorithm matching~\cite{AlmanW2015}.

A number of studies address approximate versions of inner-product counting
in subquadratic time, such as the detection of outlier correlations and 
offline computation of approximate nearest neighbors, including
Valiant~\cite{Valiant2015},
Karppa, Kaski, and Kohonen~\cite{KarppaKK2018},
Alman~\cite{Alman2019}, and
Alman, Chan, and Williams~\cite{AlmanCW2020}.
All the algorithms above utilize fast rectangular matrix multiplication.

{\em Permanents.}
Bax and Franklin presented a randomised $2^{n-\Omega(n^{1/3}/\log n)}$ expected time algorithm for the $0/1$-matrix permanent~\cite{BaxF2002}.
Bj\"orklund~\cite{Bjorklund2016} derived a faster and deterministic $2^{n-\Omega(\sqrt{n/\log n})}$ time algorithm.
The algorithm was subsequently improved to a deterministic $2^{n-\Omega(\sqrt{n/\log \log n})}$ time algorithm by Bj\"orklund, Kaski, and Williams~\cite{BjorklundKW2019}.

For the computation of an integer matrix permanent modulo a prime power $p^{\lambda n/p}$ for any constant $\lambda<1$, Bj\"orklund, Husfeldt, and Lyckberg~\cite{BjorklundHL2017} derived a $2^{n-\Omega(n/(p\log p))}$ time algorithm. For the computation of a matrix permanent over an arbitrary ring $R$ on $r$ elements, Bj\"orklund and Williams~\cite{BjorklundW2019} gave a deterministic $2^{n-\Omega(\frac{n}{r})}$ time algorithm.

The problem \#\textsc{InnerProduct} has various applications in combinatorial 
algorithms. To mention two in particular, it can be used to count the 
satisfying assignments to a
 \textsc{Sym}$\circ$\textsc{And} formula (cf.~Sect.~\ref{sect:sym-and}), 
or compute the weight enumerator polynomial of a linear code 
(cf.~Sect.~\ref{sect:weight-enum}).

\section{Reconstruction from Sum-Aggregates by Prime Residue}

\label{sect:reconstruction}

This section develops the main methodological contribution of this work. 
Namely, we show that a complex-valued function 
$f:D\rightarrow\mathbb{C}$ can be reconstructed from its 
sum-aggregates by prime residue when the domain $D$ is a prefix of the 
set of nonnegative integers. In essence, reconstruction of a function from 
its sum-aggregates can be viewed as an {\em additive} analog of 
the Chinese Remainder Theorem; that is, we obtain reconstruction
up to the {\em sum} of the prime moduli---in the precise sense 
of~\eqref{eq:sm} below---whereas the Chinese Remainder Theorem enables 
reconstruction up to the product of the moduli.%
\footnote{%
Here it should be noted that the scope of the Chinese Remainder Theorem is
also somewhat more restricted than our present setting; indeed, in our setting
the Chinese Remainder Theorem does not enable the reconstruction of 
an arbitrary function $f$ but rather is restricted to reconstruction in 
the case when $f$ is known to vanish in all but one point of $D$.}

In our application of counting pairs of vectors by inner product, we let $f$ 
be a counting function such that $f(\ell)$ counts the number of pairs 
$(x,y)\in\mathcal{A}\times\mathcal{B}$ with $\bra x,y\ket=\ell$. Reconstruction 
from sum-aggregates then enables us to recover $f$ by counting the number of 
pairs $(x,y)$ with $\bra x,y\ket\equiv r\pmod p$ for small primes $p$ and residues 
$r\in\{0,1,\ldots,p-1\}$; we postpone the details of this application to 
Sect.~\ref{sect:main-proof} and first proceed to study reconstructibility. 

\subsection{Sum-Aggregation by Prime Residue}
Let $p_1,p_2,\ldots,p_m$ be distinct prime numbers and let us assume that
\[
D\subseteq\bigl\{0,1,\ldots,s_m-1\bigr\}
\]
where
\begin{equation}
\label{eq:sm}
s_m=1+\sum_{b=1}^m\bigl(p_b-1\bigr)\,.
\end{equation} Letting $f_{\ell}$ be shorthand for $f(\ell)$, we show that we can recover $f$ from the sequence of its {\em sum-aggregates}
\begin{equation}
\label{eq:F}
F_{br}=\!\!\!\sum_{\substack{\ell\in D\\\ell\equiv r\!\!\!\!\pmod {p_b}}}\!\!\! f_\ell
\end{equation}
for each residue $r\in\{0,1,\ldots,p_b-1\}$ and each $b\in\{1,2,\ldots,m\}$. 

To start with, let us observe that this sequence is linearly redundant.
Indeed, define the sum
\begin{equation}
\label{eq:F0}
F_{01}=\sum_{\ell\in D} f_\ell
\end{equation}
and observe that for each $b\in\{1,2,\ldots,m\}$ we have the linear relation
\[
F_{01}=\sum_{r=0}^{p_b-1} F_{br}\,.
\]
To obtain an equivalent and---as we will shortly show---linearly 
irredundant sequence, take the sequence formed by the sum $F_{01}$ 
followed by $F_{br}$ for each {\em nonzero} residue 
$r\in\{1,2,\ldots,p_b-1\}$ and each $b\in\{1,2,\ldots,m\}$.
Let us write $F$ for this sequence of length $s_m$.
By extending the domain of the function $f$ with zero-values $f_\ell=0$
as needed, we can also assume that $D=\{0,1,\ldots,s_m-1\}$ in what follows.

\subsection{Sum-Aggregation as a Linear System}
Let us now study reconstruction of $f$ from $F$. From \eqref{eq:F} and
\eqref{eq:F0} we observe that the task of reconstructing $f$ from $F$ 
is equivalent to solving the linear system 
\begin{equation}
\label{eq:FAf}
F=Af\,,
\end{equation}
where $A$ is the $s_m\times s_m$ {\em nonzero residue aggregation matrix} 
whose entries are defined for all $b\in\{0,1,2,\ldots,m\}$, 
$i\in\{1,2,\ldots,p_b-1\}$, and 
$\ell\in\{0,1,\ldots,s_m-1\}$ by the rule
\begin{equation}
\label{eq:A-entry}
A_{bi,\ell}=
\begin{cases}
1 & \text{if $b=0$;}\\
1 & \text{if $b\geq 1$ and $i\equiv \ell\!\!\pmod{p_b}$;}\\
0 & \text{if $b\geq 1$ and $i\not\equiv \ell\!\!\pmod{p_b}$,}
\end{cases}
\end{equation}
where we have assumed for convenience that $p_0=2$.
Indeed, 
we readily verify from \eqref{eq:F}, \eqref{eq:F0}, and~\eqref{eq:A-entry} 
that
\[
F_{bi}=\sum_{\ell=0}^{s_m-1} A_{bi,\ell}f_\ell
\]
holds for each $b\in\{0,1,\ldots,m\}$ and $i\in\{0,1,\ldots,p_b-1\}$.
When we want to stress the $m$ selected primes, 
we write $A^{p_1,p_2,\ldots,p_m}$ for the matrix $A$.

The row-banded structure given by \eqref{eq:A-entry} is perhaps easiest 
illustrated with a small example. Below we display the matrix $A$ for 
the primes $p_1=2$, $p_2=3$, and $p_3=5$:
\[
A^{2,3,5}=
\left[\begin{array}{cccccccc}
1&1&1&1&1&1&1&1\\\hline
0&1&0&1&0&1&0&1\\\hline
0&1&0&0&1&0&0&1\\
0&0&1&0&0&1&0&0\\\hline
0&1&0&0&0&0&1&0\\
0&0&1&0&0&0&0&1\\
0&0&0&1&0&0&0&0\\
0&0&0&0&1&0&0&0
\end{array}\right]\,.
\]
Observe in particular that the first band $b=0$ 
corresponds to the sum~\eqref{eq:F0} and the subsequent bands 
$b\in\{1,2,\ldots,m\}$ 
each correspond to one of the primes $p_1,p_2,\ldots,p_m$ so that 
the $p_b-1$ rows inside each band correspond to the sum-aggregates \eqref{eq:F}
of the $p_b-1$ {\em nonzero} residue classes modulo~$p_b$.

Our main technical lemma establishes that the matrix $A$ is invertible,
thus enabling reconstruction of $f$ from $F$.

\begin{Lem}[Reconstruction from Sum-Aggregates by Prime Residue]
\label{lem:reconstruction}
The nonzero residue aggregation matrix $A^{p_1,p_2,\ldots,p_m}$ 
is invertible whenever $p_1,p_2,\ldots,p_m$ are distinct primes.
\end{Lem}
The key idea in the proof is
to decompose $A^{p_1,p_2,\ldots,p_m}$ over the complex numbers into the
product of a near-block-diagonal matrix with near-Vandermonde blocks
and a Vandermonde matrix, both of which are then shown to have nonzero 
determinant. The rest of this section is devoted to a proof of
Lemma~\ref{lem:reconstruction}.

\subsection{Preliminaries on Complex Roots of Unity}

We will need the following standard facts about complex roots of unity.
For a positive integer $N$, let us write 
\[
\omega_{N}=\exp\biggl(\frac{2\pi\Im}{N}\biggr)\,,
\]
where $\Im=\sqrt{-1}$ is the imaginary unit. For all $m\in\mathbb{Z}$ we have
\begin{equation}
\label{eq:root-power-sum}
\frac{1}{N}\sum_{j=0}^{N-1}\omega_{N}^{km}=
\begin{cases}
1 & \text{if $k\equiv 0\!\!\pmod N$};\\
0 & \text{if $k\not\equiv 0\!\!\pmod N$}.
\end{cases}
\end{equation}

\subsection{Reconstruction from Sum-Aggregates---Proof of Lemma~\ref{lem:reconstruction}}

We show that for distinct primes $p_1,p_2,\ldots,p_m$ 
the matrix $A=A^{p_1,p_2,\ldots,p_m}$ is invertible over rational numbers.
Our strategy is to 
show that $A=UV$ for two complex matrices $U$ and $V$ that both have nonzero
determinant. Indeed, the near-cyclic banded structure of $A$ suggests that one 
should pursue a decomposition in terms of block-structured near-Vandermonde
matrices. Let us first define the matrices $U$ and $V$, then present 
a small example, and then complete the proof.

The matrix $U$ will use a $(m+1)\times(m+1)$ block structure that is similar
to the $(m+1)$-band structure of $A$, but now the structure is used both for
rows and columns. Again for convenience we assume $p_0=2$. The matrix $U$ is  
defined for all $b\in\{0,1,2,\ldots,m\}$, 
$i\in\{1,2,\ldots,p_b-1\}$, 
$d\in\{0,1,\ldots,m\}$,
and 
$k\in\{1,2,\ldots,p_d-1\}$ by the rule
\begin{equation}
\label{eq:U-entry}
U_{bi,dk}=
\begin{cases}
1 & \text{if $d=0$ and $b=0$;}\\
\frac{1}{p_b} & \text{if $d=0$ and $b\geq 1$;}\\
0 & \text{if $d\geq 1$ and $b\neq d$;}\\
\frac{1}{p_b} \omega_{p_b}^{-ik} & \text{if $d\geq 1$ and $b=d$.}\\
\end{cases}
\end{equation}
The matrix $V$ is a Vandermonde matrix with $(m+1)$-banded structure
defined for all $d\in\{0,1,\ldots,m\}$, 
$k\in\{1,2,\ldots,p_d-1\}$, and
$\ell\in\{0,1,\ldots,s_m-1\}$ by the rule
\begin{equation}
\label{eq:V-entry}
V_{dk,\ell}=
\begin{cases}
1                    & \text{if $d=0$;}\\
\omega_{p_d}^{k\ell} & \text{if $d\geq 1$}\,.
\end{cases}
\end{equation}
Before proceeding with the proof that $A=UV$, let us present an example 
for the primes $p_1=2$, $p_2=3$, and $p_3=5$. We have
\begin{equation}
\label{eq:AUV-example}
\tiny
\begin{split}
\hspace*{1cm}&\hspace*{-1cm}\left[\begin{array}{cccccccc}
1&1&1&1&1&1&1&1\\\hline
0&1&0&1&0&1&0&1\\\hline
0&1&0&0&1&0&0&1\\
0&0&1&0&0&1&0&0\\\hline
0&1&0&0&0&0&1&0\\
0&0&1&0&0&0&0&1\\
0&0&0&1&0&0&0&0\\
0&0&0&0&1&0&0&0
\end{array}\right]_{A^{2,3,5}}
=\\[5mm]
&=
\left[\begin{array}{c|c|cc|cccc}
1&0&0&0&0&0&0&0\\\hline
\frac{1}{2}&\frac{1}{2}\omega_2^{-1\cdot 1}&0&0&0&0&0&0\\\hline
\frac{1}{3}&0&\frac{1}{3}\omega_3^{-1\cdot 1}&\frac{1}{3}\omega_3^{-1\cdot 2}&0&0&0&0\\
\frac{1}{3}&0&\frac{1}{3}\omega_3^{-2\cdot 1}&\frac{1}{3}\omega_3^{-2\cdot 2}&0&0&0&0\\\hline
\frac{1}{5}&0&0&0&\frac{1}{5}\omega_5^{-1\cdot 1}&\frac{1}{5}\omega_5^{-1\cdot 2}&\frac{1}{5}\omega_5^{-1\cdot 3}&\frac{1}{5}\omega_5^{-1\cdot 4}\\
\frac{1}{5}&0&0&0&\frac{1}{5}\omega_5^{-2\cdot 1}&\frac{1}{5}\omega_5^{-2\cdot 2}&\frac{1}{5}\omega_5^{-2\cdot 3}&\frac{1}{5}\omega_5^{-2\cdot 4}\\
\frac{1}{5}&0&0&0&\frac{1}{5}\omega_5^{-3\cdot 1}&\frac{1}{5}\omega_5^{-3\cdot 2}&\frac{1}{5}\omega_5^{-3\cdot 3}&\frac{1}{5}\omega_5^{-3\cdot 4}\\
\frac{1}{5}&0&0&0&\frac{1}{5}\omega_5^{-4\cdot 1}&\frac{1}{5}\omega_5^{-4\cdot 2}&\frac{1}{5}\omega_5^{-4\cdot 3}&\frac{1}{5}\omega_5^{-4\cdot 4}
\end{array}\right]_{U^{2,3,5}}\cdot\\[5mm]
&\ \,\,\cdot
\left[\begin{array}{cccccccc}
1&1&1&1&1&1&1&1\\\hline
\omega_2^{1\cdot 0}&\omega_2^{1\cdot 1}&\omega_2^{1\cdot 2}&\omega_2^{1\cdot 3}&\omega_2^{1\cdot 4}&\omega_2^{1\cdot 5}&\omega_2^{1\cdot 6}&\omega_2^{1\cdot 7}\\\hline
\omega_3^{1\cdot 0}&\omega_3^{1\cdot 1}&\omega_3^{1\cdot 2}&\omega_3^{1\cdot 3}&\omega_3^{1\cdot 4}&\omega_3^{1\cdot 5}&\omega_3^{1\cdot 6}&\omega_3^{1\cdot 7}\\
\omega_3^{2\cdot 0}&\omega_3^{2\cdot 1}&\omega_3^{2\cdot 2}&\omega_3^{2\cdot 3}&\omega_3^{2\cdot 4}&\omega_3^{2\cdot 5}&\omega_3^{2\cdot 6}&\omega_3^{2\cdot 7}\\\hline
\omega_5^{1\cdot 0}&\omega_5^{1\cdot 1}&\omega_5^{1\cdot 2}&\omega_5^{1\cdot 3}&\omega_5^{1\cdot 4}&\omega_5^{1\cdot 5}&\omega_5^{1\cdot 6}&\omega_5^{1\cdot 7}\\
\omega_5^{2\cdot 0}&\omega_5^{2\cdot 1}&\omega_5^{2\cdot 2}&\omega_5^{2\cdot 3}&\omega_5^{2\cdot 4}&\omega_5^{2\cdot 5}&\omega_5^{2\cdot 6}&\omega_5^{2\cdot 7}\\
\omega_5^{3\cdot 0}&\omega_5^{3\cdot 1}&\omega_5^{3\cdot 2}&\omega_5^{3\cdot 3}&\omega_5^{3\cdot 4}&\omega_5^{3\cdot 5}&\omega_5^{3\cdot 6}&\omega_5^{3\cdot 7}\\
\omega_5^{4\cdot 0}&\omega_5^{4\cdot 1}&\omega_5^{4\cdot 2}&\omega_5^{4\cdot 3}&\omega_5^{4\cdot 4}&\omega_5^{4\cdot 5}&\omega_5^{4\cdot 6}&\omega_5^{4\cdot 7}\\
\end{array}\right]_{V^{2,3,5}}\,.
\end{split}
\end{equation}
The main technical aspect of the proof that $A=UV$ is to partition
the index $\ell\in\{0,1,\ldots,s_m-1\}$ to the $m+1$ bands. 
Towards this end, define for each $c\in\{0,1,\ldots,m\}$ the prefix-sum
\[
\underline{s}_c=
\begin{cases}
0                                        & \text{if $c=0$;}\\
1 + \sum_{\ell=1}^{c-1}\bigl(p_c-1\bigr) & \text{if $c\geq 1$.}
\end{cases}
\]
In particular, 
for every $\ell\in\{0,1,\ldots,s_m-1\}$, we observe that there exist unique
$c\in\{0,1,\ldots,m\}$ and $j\in\{1,2,\ldots,p_c-1\}$ such that 
\begin{equation}
\label{eq:ell-decomposition}
\ell=j-1+\underline{s}_c\,.
\end{equation}

We are now ready to show that $A=UV$.
Let $b\in\{0,1,\ldots,m\}$, $i\in\{0,1,\ldots,p_b-1\}$,
and $\ell\in\{0,1,\ldots,s_k-1\}$ be arbitrary. 
Let $c\in\{0,1,\ldots,m\}$ and $j\in\{1,2,\ldots,p_c-1\}$ be uniquely 
determined from $\ell$ by \eqref{eq:ell-decomposition}.
From \eqref{eq:U-entry}, \eqref{eq:V-entry}, 
\eqref{eq:root-power-sum}, and \eqref{eq:A-entry} we observe that 
\[
\small
\begin{split}
\,\,\,&\!\!\!\sum_{d=0}^m\sum_{k=0}^{p_d-1} U_{bi,dk}V_{dk,\ell}
=\\[2mm]
&=
\begin{cases}
1 & \!\!\!\text{if $b=0$};\\[1mm]
\frac{1}{p_b}\bigl(1+\sum_{k=1}^{p_b-1}\omega_{p_b}^{-ik+k(j-1+\underline{s}_b)}\bigr)=\frac{1}{p_b}\sum_{k=0}^{p_b-1}\omega_{p_b}^{k(j-i-1+\underline{s}_b)}=1 & \!\!\!\text{if $b\geq 1$ and $i\equiv j-1+\underline{s}_b=\ell\!\!\!\!\pmod{p_b}$};\\[1mm]
\frac{1}{p_b}\bigl(1+\sum_{k=1}^{p_b-1}\omega_{p_b}^{-ik+k(j-1+\underline{s}_b)}\bigr)=\frac{1}{p_b}\sum_{k=0}^{p_b-1}\omega_{p_b}^{k(j-i-1+\underline{s}_b)}=0 & \!\!\!\text{if $b\geq 1$ and $i\not\equiv j-1+\underline{s}_b=\ell\!\!\!\!\pmod{p_b}$.}
\end{cases}\\[2mm]
&=A_{bi,\ell}\,.
\end{split}
\]
Thus, $A=UV$ holds.
It remains to show that both matrices $U$ and $V$ have nonzero determinant
over the complex numbers. Starting with the Vandermonde matrix $V$,
let $\nu_0=1$ and $\nu_\ell=\omega_{p_c}^j$ for $\ell\in\{1,2,\ldots,s_m-1\}$,
where $c\in\{0,1,\ldots,m\}$ and $j\in\{1,2,\ldots,p_c-1\}$ are uniquely 
determined from $\ell$ by \eqref{eq:ell-decomposition}. In particular,
we observe that $V$ is a Vandermonde matrix with $D=s_m-1$ and 
\[
V=
\left[\begin{array}{cccc}
\nu_0^0 & \nu_0^1 & \cdots & \nu_0^D\\
\nu_1^0 & \nu_1^1 & \cdots & \nu_1^D\\
\vdots & \vdots & & \vdots \\
\nu_D^0 & \nu_D^1 & \cdots & \nu_D^D\\
\end{array}\right]\,.
\]
The Vandermonde determinant formula thus gives 
\[
\det V=\sum_{0\leq k<\ell\leq D}(\nu_\ell-\nu_k)\,.
\]
Furthermore, this determinant is nonzero because $p_1,p_2,\ldots,p_m$ are 
distinct primes and thus $\nu_0,\nu_1,\ldots,\nu_D$ are distinct.
Next, let us consider the matrix $U$ defined by \eqref{eq:U-entry}. 
At this point it may be useful to revisit the structure of $U$ via 
the example \eqref{eq:AUV-example}.
We observe that the block-diagonal of $U$ with
$b=c\geq 1$ consists of matrices that each decompose into the product of a 
$(p_b-1)\times (p_b-1)$ diagonal matrix
with diagonal entries $\frac{1}{p_b}\omega_{p_b}^{-i}$ for 
$i\in\{1,2,\ldots,p_b-1\}$ and a $(p_b-1)\times (p_b-1)$ Vandermonde matrix
with a nonzero determinant since $\omega_{p_b}^{-i}$ 
for $i\in\{1,2,\ldots,p_b-1\}$ are distinct. Thus, since the determinant
of $U$ is the product of the determinants of the block-matrices on 
the diagonal, each of which is nonzero, the determinant of $U$ is nonzero.
It follows that $A$ is invertible and thus given $F$ we can
solve for $f$ via \eqref{eq:FAf}.
This completes the proof of Lemma~\ref{lem:reconstruction}. $\qed$

\section{Counting Pairs of Zero-One Vectors by Inner Product}

\label{sect:main-proof}

This section documents our main algorithm and proves Theorem~\ref{thm:main}.
Let $\kappa$ be a parameter that satisfies, with foresight,
\begin{equation}
\label{eq:kappa-up}
4\leq\kappa\leq\frac{\log n}{(\log\log n)^4}\,.
\end{equation}
Let $a^{(1)},a^{(2)},\ldots,a^{(n)}\in\{0,1\}^d$ and 
$b^{(1)},b^{(2)},\ldots,b^{(n)}\in\{0,1\}^d$ be given as input 
with $d\leq\kappa\log n$. 
We want to compute for each $t\in\{0,1,\ldots,d\}$ the count
\[
f_t=|\{(i,j)\in\{1,2,\ldots,n\}^2:\bra a^{(i)},b^{(j)}\ket=t\}|\,.
\]

Our high-level approach will be to use Lemma~\ref{lem:reconstruction} and
\eqref{eq:FAf} to solve for the counts $f_0,f_1,\ldots,f_d$ using as
input counts that have been sum-aggregated by prime residue. 
More precisely, we will work with prime moduli $p_1,p_2,\ldots,p_m$ 
and develop an algorithm that computes, for given 
further input $p\in\{p_1,p_2,\ldots,p_m\}$ and $r\in\{0,1,\ldots,p-1\}$,
the sum-aggregated count
\[
F_{pr}=|\{(i,j)\in\{1,2,\ldots,n\}^2:\bra a^{(i)},b^{(j)}\ket\equiv r\!\!\!\pmod p\}|\,.
\]
The detailed choices for $m$ and the primes $p_1,p_2,\ldots,p_m$ will be 
presented later.

\subsection{The Residue-Indicator Polynomial}
Assume $p$ and $r$ have been given. We will rely on the polynomial 
method, and accordingly we first build a standard polynomial that 
indicates the residue $r$ modulo $p$ in a pair of vectors.

Let $x=(x_1,x_2,\ldots,x_d)$ and $y=(y_1,y_2,\ldots,y_d)$ be two
vectors of indeterminates. By Fermat's little theorem, 
the $2d$-indeterminate polynomial
\begin{equation}
\label{eq:indicator-poly}
G_{p,r}\bigl(x,y\bigr)
=1-\biggl(\sum_{k=1}^d x_ky_k-r\biggr)^{p-1}
\end{equation}
satisfies for all $i,j\in\{1,2,\ldots,n\}$ the indicator property
\begin{equation}
\label{eq:indicator-property}
G_{p,r}\bigl(a^{(i)},b^{(j)}\bigr)
\equiv
\begin{cases}
1\!\!\!\pmod p & \text{if $\bra a^{(i)},b^{(j)}\ket\equiv r\!\!\!\pmod p$;}\\
0\!\!\!\pmod p & \text{if $\bra a^{(i)},b^{(j)}\ket\not\equiv r\!\!\!\pmod p$}.
\end{cases}
\end{equation}
We observe that $G_{p,r}$ has degree $2p-2$.

\subsection{Modulus Amplification for Zero-One Residues}

To enable taking the sum of a large number of indicators, we make use 
of the {\em modulus amplifying polynomials} of Beigel and 
Tarui~\cite{BeigelT1994}.

\begin{Thm}[Modulus amplification; Beigel and Tarui~\cite{BeigelT1994}]
\label{thm:amp}
For $h\in\mathbb{Z}_{\geq 1}$, define the polynomial
\begin{equation}
\label{eq:modamp-poly}
A_h(z)=1-(1-z)^h\sum_{j=0}^{h-1}\binom{h+j-1}{j}z^j\,.
\end{equation}
Then, for all $m\in\mathbb{Z}_{\geq 2}$ and $s\in\mathbb{Z}$, we have 
\begin{enumerate}
\item[(i)]
$s\equiv 0\pmod m$ implies $A_h(s)\equiv 0\pmod{m^h}$, and 
\item[(ii)]
$s\equiv 1\pmod m$ implies $A_h(s)\equiv 1\pmod{m^h}$.
\end{enumerate}
\end{Thm}

We observe that $A_h$ has degree $2h-1$. 
Composing \eqref{eq:modamp-poly} and \eqref{eq:indicator-poly}, we obtain the
amplified residue-indicator polynomial
\begin{equation}
\label{eq:amp-indicator-poly}
G_{p,r}^h(x,y)=A_h\bigl(G_{p,r}(x,y)\bigr)\,.
\end{equation}
From \eqref{eq:indicator-property} and Theorem~\ref{thm:amp}, we observe
the amplified indicator property
\begin{equation}
\label{eq:amp-indicator-property}
G_{p,r}^h\bigl(a^{(i)},b^{(j)}\bigr)
\equiv
\begin{cases}
1\!\!\!\pmod{p^h} & \text{if $\bra a^{(i)},b^{(j)}\ket\equiv r\!\!\!\pmod p$;}\\
0\!\!\!\pmod{p^h} & \text{if $\bra a^{(i)},b^{(j)}\ket\not\equiv r\!\!\!\pmod p$}.
\end{cases}
\end{equation}
Furthermore, we observe that $G_{p,r}^h$ has degree $(2h-1)(2p-2)$.

\subsection{Multilinear Reduct and Bounding the Number of Monomials}
For a nonnegative integer $e$, define $\underline{e}=0$ if $e=0$ 
and $\underline{e}=1$ if $e\geq 1$. For a monomial 
$x_1^{e_1}x_2^{e_2}\cdots x_d^{e_d}y_1^{f_1}y_2^{f_2}\cdots y_d^{f_d}$,
define the {\em multilinear reduct} by
\[
\underline{x_1^{e_1}x_2^{e_2}\cdots x_d^{e_d}y_1^{f_1}y_2^{f_2}\cdots y_d^{f_d}}
=
x_1^{\underline{e}_1}x_2^{\underline{e}_2}\cdots x_d^{\underline{e}_d}y_1^{\underline{f}_1}y_2^{\underline{f}_2}\cdots y_d^{\underline{f}_d}\,.
\]
For a polynomial $Q(x,y)$, define the multilinear reduct $\underline{Q}(x,y)$ 
by taking the multilinear reduct of each monomial $Q(x,y)$ and simplifying.
Since $a^{(i)}$ and $b^{(j)}$ are $\{0,1\}$-valued vectors, over the
integers we have
\begin{equation}
\label{eq:multilin-reduct}
Q\bigl(a^{(i)},b^{(j)}\bigr)=\underline{Q}\bigl(a^{(i)},b^{(j)}\bigr)\,.
\end{equation}
Furthermore, if $Q$ has degree $D$, then $\underline{Q}$ has at most
$\sum_{j=0}^D\binom{2d}{j}$ monomials. In particular, we observe that
$\underline{G}_{p,r}^h$ has at most $\sum_{j=0}^{4hp}\binom{2d}{j}$ monomials.

\subsection{Split-Monomial Form of the Multilinear Reduct}
Suppose that the multilinear reduct $\underline{G}_{p,r}^h(x,y)$ has 
exactly $M$ monomials with the representation
\begin{equation}
\label{eq:monomial-rep}
\underline{G}_{p,r}^h(x,y)=
\sum_{k=1}^M 
\gamma^{(k)}\,
x_1^{e_1^{(k)}}\!\!x_2^{e_2^{(k)}}\!\!\cdots x_d^{e_d^{(k)}}\!\!
y_1^{f_1^{(k)}}\!\!y_2^{f_2^{(k)}}\!\!\cdots y_d^{f_d^{(k)}}\,.
\end{equation}
For $I,J\subseteq\{1,2,\ldots,n\}$ and $k\in\{1,2,\ldots,M\}$, define
\begin{equation}
\label{eq:lr}
L_{I,k}=
\sum_{i\in I}
\bigl(a_1^{(i)}\bigr)^{e_1^{(k)}}\!\!\bigl(a_2^{(i)}\bigr)^{e_2^{(k)}}\!\!\cdots \bigl(a_d^{(i)}\bigr)^{e_d^{(k)}}\gamma^{(k)}\,,\qquad
R_{J,k}=
\sum_{j\in J}
\bigl(b_1^{(j)}\bigr)^{f_1^{(k)}}\!\!\bigl(b_2^{(j)}\bigr)^{f_2^{(k)}}\!\!\cdots \bigl(b_d^{(j)}\bigr)^{f_d^{(k)}}\,.
\end{equation}
From \eqref{eq:lr}, \eqref{eq:monomial-rep}, \eqref{eq:multilin-reduct},
and \eqref{eq:amp-indicator-property}, we have
\begin{equation}
\label{eq:mm-count-identity}
\sum_{k=1}^M L_{I,k}R_{J,k}
=\sum_{i\in I}\sum_{j\in J}
\underline{G}_{p,r}^h\bigl(a^{(i)},b^{(j)}\bigr)
\equiv
\big|\bigl\{(i,j)\in I\times J:\bra a^{(i)},b^{(j)}\ket\equiv r\!\!\!\pmod p\bigr\}\big|\!\!\!
\pmod{p^h}\,.
\end{equation}
In particular, assuming that 
$|I||J|\leq p^h-1$, from \eqref{eq:mm-count-identity} 
it follows that 
$\sum_{k=1}^M L_{I,k}R_{J,k}$ computed modulo $p^h$ recovers 
the number of pairs $(i,j)\in I\times J$ with 
$\bra a^{(i)},b^{(j)}\ket\equiv r\pmod p$.

We now move from deriving the polynomial and its properties to 
describing the algorithm.

\subsection{Algorithm for the Prime-Residue Count}

The algorithm will rely on \eqref{eq:mm-count-identity} via 
fast rectangular matrix multiplication to count the number of 
pairs $(i,j)\in\{1,2,\ldots,n\}^2$ that satisfy 
$\bra a^{(i)},b^{(j)}\ket\equiv r\pmod p$.

The algorithm first computes the explicit $M$-monomial representation 
of the polynomial $\underline{G}^h_{p,r}$ in~\eqref{eq:monomial-rep}.
More precisely, the algorithm evaluates 
\eqref{eq:indicator-poly}, \eqref{eq:modamp-poly}, and
\eqref{eq:amp-indicator-poly} in explicit monomial representation, taking
multilinear reducts with respect to the variables
$x_1,x_2,\ldots,x_d,y_1,y_2,\ldots,y_d$ whenever possible. 
This results in the set
\begin{equation}
\label{eq:monomial-list}
\{(k,\gamma^{(k)},
     e_1^{(k)},e_2^{(k)},\ldots,e_d^{(k)},
     f_1^{(k)},f_2^{(k)},\ldots,f_d^{(k)}):k\in\{1,2,\ldots,M\}\}\,.
\end{equation}

Next, the algorithm constructs two rectangular matrices $S$ and $T$, with
the objective of making use of the following algorithm of 
Coppersmith~\cite{Coppersmith1982}.

\begin{Thm}[Coppersmith~\cite{Coppersmith1982}]
\label{thm:coppersmith}
Given an $N\times \lfloor N^{0.17}\rfloor$ matrix $S$ 
and an $\lfloor N^{0.17}\rfloor\times N$ matrix $T$ as input, 
the matrix product $ST$ over the integers can be computed in $O(N^2\log^2 N)$ 
arithmetic operations.
\end{Thm}

Towards this end, let $g$ be a positive integer whose value we will fix
later. Introduce two set partitions of $\{1,2,\ldots,n\}$ with cells 
\[
I_1,I_2,\ldots,I_{\lceil n/g\rceil}\subseteq\{1,2,\ldots,n\}\qquad
\text{and}\qquad
J_1,J_2,\ldots,J_{\lceil n/g\rceil}\subseteq\{1,2,\ldots,n\}\,,
\]
respectively, so that 
$|I_u|=g$ and $|I_v|=g$ for $u,v\in\{1,2,\ldots,\lfloor n/g\rfloor\}$. 
Indeed, we thus have 
\[
|I_u||J_v|\leq g^2
\]
for all $u,v\in\{1,2,\ldots,\lceil n/g\rceil\}$, 
so \eqref{eq:mm-count-identity} applied to $I_u$ and $J_v$ 
modulo $p^h$ recovers the number of pairs $(i,j)\in I_u\times J_v$ with 
$\bra a^{(i)},b^{(j)}\ket\equiv r\pmod p$, assuming that $g^2\leq p^h-1$,
which will be justified by our eventual choice of $g$.

Now let $N=\lceil n/g\rceil$ and define the $N\times M$
and $M\times N$ matrices $S$ and $T$ by setting, 
for $u,v\in\{1,2,\ldots,\lceil n/g\rceil\}$ and $k\in\{1,2,\ldots,M\}$, 
\[
S_{uk}=L_{I_u,k}\qquad\text{and}\qquad T_{kv}=R_{I_v,k}\,.
\]
Concretely, the algorithm computes $S$ and $T$ from the given input
one entry at a time using the computed monomial 
list \eqref{eq:monomial-list} 
and the formulas \eqref{eq:lr} for $I=I_u$ and $J=J_v$ 
for each $u,v\in\{1,2,\ldots,\lceil n/g\rceil\}$ and $k=1,2,\ldots,M$.
The algorithm then multiplies $S$ and $T$ to obtain the product 
matrix $ST$ modulo $p^h$, where we assume that each entry of $ST$ is 
reduced to $\{0,1,\ldots,p^h-1\}$. Finally, the algorithm outputs the sum
\[
F_{pr}=\sum_{u=1}^{\lceil n/g\rceil}\sum_{v=1}^{\lceil n/g\rceil}(ST)_{uv}\,.
\]

\subsection{Parameterizing the Algorithm}

Let us now start parameterizing the algorithm. First, 
to apply the algorithm in Theorem~\ref{thm:coppersmith} to the matrices 
$S$ and $T$, we need $M\leq N^{0.17}=\lceil n/g\rceil^{0.17}$. 
Subject to the assumption $g\leq n^{0.1}$---to be justified later---it 
will be sufficient to show that $M\leq n^{0.15}$.
We recall that $M\leq\sum_{j=0}^{4hp}\binom{2d}{j}$ and $d\leq \kappa\log n$.
With foresight, let us set
\begin{equation}
\label{eq:beta-kappa}
\beta_\kappa=\frac{K}{\log\kappa}\,,
\end{equation}
where $K>0$ is a small constant that will be fixed later.
In particular, since $\kappa\geq 4$, we have the upper bound
\begin{equation}
\label{eq:base-decay}
\beta_\kappa\log\frac{\kappa}{\beta_\kappa}=K-\frac{K\log K}{\kappa}+\frac{K\log\log\kappa}{\kappa}\leq \frac{5K}{4}
\end{equation}
which we can make an arbitrarily small and positive by choosing 
a small enough $K$.
Let us assume---to be justified later---that $p=o(\beta_\kappa\log n)$.
Taking 
\begin{equation}
\label{eq:h}
h=\biggl\lfloor\beta_\kappa\frac{\log n}{p}\biggr\rfloor
\end{equation}
we have, for all large enough~$n$, 
\begin{equation}
\label{eq:m-bound}
\begin{split}
M&\leq
\sum_{j=0}^{4hp}\binom{2d}{j}
\leq 
4hp\binom{2d}{4hp}
\leq 
4hp\biggl(\frac{2ed}{4hp}\biggr)^{4hp}\\
&\leq
4\biggl(\beta_\kappa\frac{\log n}{p}+1\biggr)p\biggl(\frac{2e\kappa\log n}{4\bigl(\beta_\kappa\frac{\log n}{p}-1\bigr)p}\biggr)^{4\bigl(\beta_\kappa\frac{\log n}{p}+1\bigr)p}\\
&=
4\bigl(\beta_\kappa\log n+p\bigr)\biggl(\frac{2e\kappa\log n}{4\bigl(\beta_\kappa\log n-p\bigr)}\biggr)^{4(\beta_\kappa\log n+p)}\\
&\leq
\bigl(5\beta_\kappa\log n\bigr)\biggl(\frac{2e\kappa}{3\beta_\kappa}\biggr)^{5\beta_\kappa\log n}\\
&\leq
n^{0.15}\,,
\end{split}
\end{equation}
where the last inequality follows by \eqref{eq:base-decay} and choosing $K$
small enough. 
Thus, Theorem~\ref{thm:coppersmith} applies, subject to the assumptions
$g\leq n^{0.1}$, $g^2\leq p^h-1$, and $p=o(\beta_\kappa\log n)$, which still need to 
be established. Before this, we digress to further preliminaries to enable
reconstruction. 

\subsection{Preliminaries on Asymptotics of Primes}

In what follows let us write $p_j$ for the $j$th prime number with $j=1,2,\ldots$; 
that is, $p_1=2$, $p_2=3$, $p_3=5$, and so forth.
Asymptotically, from the Prime Number Theorem we have $p_m\sim m\ln m$ 
(e.g.~Rosser and Schoenfeld~\cite{RosserS1962}), and the sum of the
first $m$ primes satisfies 
$\sum_{j=1}^m p_j\sim \frac{1}{2}m^2\ln m$ 
(cf.~Bach and Shallit~\cite{BachS1996}), where we write
$f(m)\sim g(m)$ if $\lim_{m\rightarrow\infty} \frac{f(m)}{g(m)}=1$.

When evaluated for the first $m$ primes, the reconstruction 
parameter~\eqref{eq:sm} thus satisfies
\begin{equation}
\label{eq:sm-growth}
s_m=1+\sum_{j=1}^m(p_j-1)\sim \frac{p_m^2}{2\ln p_m}\,.
\end{equation}
We are now ready to continue parameterization of the algorithm.

\subsection{Further Parameterization of the Algorithm}
Let $m$ be a positive integer whose value will be fixed shortly.
The algorithm will work with $p_1,p_2,\ldots,p_m$, the first $m$ prime numbers.
To reconstruct inner products of length-$d$ zero-one vectors over the 
integers, we need $d+1\leq s_m$, which for $d\leq \kappa\log n$ 
and \eqref{eq:sm-growth} means 
\[
\frac{p_m^2}{2\ln p_m}\sim \kappa\log n\,.
\]

From Bertrand's postulate it thus follows that choosing the least $m$ so that
\begin{equation}
\label{eq:pm-choice}
2\sqrt{\kappa(\ln n)\ln\ln n}\leq p_m\leq 4\sqrt{\kappa(\ln n)\ln\ln n}
\end{equation}
implies that we have $d+1\leq s_m$ for all large enough $n$
and thus reconstruction is feasible. The choice \eqref{eq:pm-choice}
also justifies our ealier assumption made in 
the context of \eqref{eq:h} and \eqref{eq:m-bound}
that $p_j=o(\beta_\kappa\log n)$ for all 
$j\in\{1,2,\ldots,m\}$; indeed, from \eqref{eq:kappa-up} 
and \eqref{eq:beta-kappa}, we have
\[
\begin{split}
\beta_\kappa\log n
&=\frac{K\log n}{\log\kappa}
\end{split}
\]
and thus from \eqref{eq:kappa-up} and \eqref{eq:pm-choice} we observe that
\[
\frac{p_j}{\beta_\kappa\log n}
\leq 
\frac{4\kappa^{1/2}(\log\kappa)(\ln n)^{1/2}(\ln\ln n)^{1/2}}{K\log n}=o(1)\,.
\]

Let us next choose the parameter $g$. Using $p_j=o(\beta_\kappa\log n)$ again, 
we have
\[
p_j^{h_j}
=p_j^{\bigl\lfloor\beta_\kappa\frac{\log n}{p_j}\bigr\rfloor}
\geq p_j^{\beta_\kappa\frac{\log n}{p_j}-1}
\geq p_j^{\beta_\kappa\frac{\log n}{2p_j}}
=2^{\beta_\kappa\frac{\log n}{2p_j}\log p_j}
=n^{\beta_\kappa\frac{\log p_j}{2p_j}}\,.
\]
Since $p_1<p_2<\cdots<p_m$, for $j\in\{1,2,\ldots,m\}$ thus
\[
p_j^{h_j} \geq n^{\beta_\kappa\frac{\log p_m}{2p_m}}\,.
\]
It follows that choosing
\begin{equation}
\label{eq:g-choice}
g
=\biggl\lfloor \sqrt{n^{\beta_\kappa\frac{\log p_m}{2p_m}}-1}\biggr\rfloor
\end{equation}
justifies our assumption $g^2\leq p_j^{h_j}-1$ for $j\in\{1,2,\ldots,m\}$.
The final assumption $g\leq n^{0.1}$ is justified by observing that
$\frac{\log p_m}{2p_m}$ is a decreasing function of $m$ and 
observing that $\beta_\kappa=o(1)$ by \eqref{eq:kappa-up} and 
\eqref{eq:beta-kappa}.

The algorithm is now parameterized. Let us proceed to analyse its running
time.

\subsection{Running Time Analysis}
First, let us seek control on $N$ as a function of $n$.
From \eqref{eq:pm-choice} and \eqref{eq:g-choice}, we have
\[
g\geq 
\sqrt{n^{\beta_\kappa\frac{2\log 2+\log\kappa+\log\ln n+\log\ln\ln n}{16\sqrt{\kappa\ln n\ln\ln n}}}-1}-1\,.
\]
This together with \eqref{eq:kappa-up} gives us the crude lower bound
\[
g=\exp\biggl(\Omega\biggl(\beta_\kappa\sqrt{\frac{(\ln n)\ln\ln n}{\kappa}}\biggr)\biggr)\,.
\]
We thus have
\[
N^2=\lceil n/g\rceil^2=n^{2-\Omega\bigl(\beta_\kappa\sqrt{\frac{\ln\ln n}{\kappa\ln n}}\bigr)}\,.
\]
Recalling \eqref{eq:m-bound}, we observe that 
the time to compute the $M$-monomial list \eqref{eq:monomial-list} can 
be bounded by $n^{0.31}$ because the algorithm is careful to take multilinear
reducts and thus at no stage of evaluating 
\eqref{eq:indicator-poly}, \eqref{eq:modamp-poly}, and
\eqref{eq:amp-indicator-poly} the number of monomials increases above
$(n^{0.15})^2=n^{0.30}$. 
Since 
\[
\log p_j^{h_j}=h_j\log p_j
=\biggl\lfloor\beta_\kappa\frac{\log n}{p_j}\biggr\rfloor\log p_j=O(\log n)\,,
\]
the arithmetic over the integers and modulo $p_j^{h_j}$ for each 
$j=1,2,\ldots,m$ runs in time polylogarithmic in $n$ for each arithmetic 
operation executed by the algorithm. 
Because the algorithm in Theorem~\ref{thm:coppersmith} runs in
$O(N^2\log^2 N)$ arithmetic operations, we observe that the polylogarithmic
terms are subsumed by the asymptotic notation and  
the entire algorithm for computing $F_{pr}$ 
for given $p\in\{p_1,p_2,\ldots,p_m\}$ and $r\in\{0,1,\ldots,p-1\}$ 
runs in time
\begin{equation}
\label{eq:runtime}
n^{2-\Omega\bigl(\beta_\kappa\sqrt{\frac{\log\log n}{\kappa\log n}}\bigr)}
=n^{2-\Omega\bigl(\sqrt{\frac{\log\log n}{\kappa(\log\kappa)^2\log n}}\bigr)}\,.
\end{equation}

From \eqref{eq:pm-choice} we observe that the required repeats for different 
$p$ and $r$ result in multiplicative polylogarithmic terms in $n$ and 
are similarly subsumed to result in total running time of the 
form~\eqref{eq:runtime}. This completes the proof of Theorem~\ref{thm:main}.
$\qed$

\section{A Faster Randomized Algorithm for \#\textsc{InnerProduct}}

\label{sect:proof-of-randomized-algorithm}

This section sketches a proof for Theorem~\ref{thm:random}.
We follow the algorithm outlined in Alman and Williams~\cite{AlmanW2015}. We note that by their Theorem~1.2, there are probabilistic polynomials over any field with error $\epsilon$ of degree $O(\sqrt{n\log(1/\epsilon)})$. In their Theorem~4.2, they have a probabilistic OR-construction that takes the disjunction of a random set of $s^2$ pairs of vector inner products
as
\[
q(x_1,y_1,x_2,y_2,\ldots,x_s,y_s)=1+\prod_{k=1}^2 \bigl(1+\sum_{(i,j)\in R_k} \bigl(1+p(x_{i,1}+y_{i,1},x_{i,2}+y_{j,2},\ldots,x_{i,s}+y_{j,s})\bigr) \bigr)\,,
\]
where $p$ is a probabilistic threshold polynomial over $\mathbb{F}_2$ of error $\epsilon=s^{-3}$, and $R_k\subseteq [s]^2$ for $k=1,2$ are sieve subsets drawn uniformly at random. 
This construction can be used to detect w.h.p. if there is a pair in the $s^2$-sized batch whose difference Hamming weight is less than the threshold.
By repeated computations with new $p$'s and $R_k$'s, a majority vote for the batch can be chosen as the correct answer, again w.h.p. for all batches.

We implement the following change of $q$ to get an \#\textsc{InnerProduct} algorithm. We take $p$ to be a probabilistic polynomial of error $\epsilon=s^{-3}$ for the symmetric function $\iv{\sum_{i=1}^n z_i=t}$, over a field of characteristic $>s^2$. We then construct $q$ as
\begin{equation}
\label{eq:newq}
q(x_1,y_1,x_2,y_2,\ldots,x_s,y_s)=\sum_{(i,j)\in[s]^2} p(x_{i,1}y_{i,1},x_{i,2}y_{j,2},\ldots,x_{i,s}y_{j,s})\,.
\end{equation}
Since the characteristic of the field is large enough, \eqref{eq:newq} is equal to the number of pairs in the $s^2$-sized batch that has inner product equal to $t$ with probability at least
$1-s^2\epsilon\geq 1-\frac{1}{s}$, a similar bound on the probability as in Theorem~4.2. Also, the degree of the polynomials is only a factor $2$ larger. As with the original algorithm,
if we repeat this enough times and take the majority in each batch, we get the correct number of pairs with $t$ as inner product in all batches. By summing these final majority numbers over the integers, we obtain the output. We note that the parameters of the error and the degree has only changed by a constant, and hence that all calculations of the running time and the error bound of the original algorithm carries through also for our modification of the algorithm. This completes the proof sketch. $\qed$

\section{A Lower Bound for \#\textsc{InnerProduct} via Zero-One Permanents}
\label{sect:perm}

This section proves Theorem~\ref{thm:perm-to-exactip}; 
the proof of Theorem~\ref{thm:perm-to-ov} is presented in 
Appendix~\ref{sect:perm-cont}.

Throughout this section we let 
$M$ be an $n\times n$ matrix with entries $m_{ij}\in\{0,1\}$ 
for $i,j\in\{1,2,\ldots,n\}$.
For convenience, let us write $[n]=\{1,2,\ldots,n\}$.
Recalling Ryser's formula, we have
\begin{equation}
\label{eq:ryser-recalled}
\per M=(-1)^n\sum_{S\subseteq[n]}(-1)^{|S|}\prod_{i\in[n]}\sum_{j\in S}m_{ij}\,.
\end{equation}

\subsection{First Reduction: Chinese Remaindering}

Since it is immediate that $0\leq\per M\leq n!$, it suffices to compute 
the permanent modulo small primes $p$ and then assemble the result over 
the integers via the Chinese Remainder Theorem. Let us first state and prove 
a crude upper bound on the size of the primes needed. 
For a positive integer $m$, let us write $m\#$ for the product of all 
prime numbers at most $m$.
\begin{Lem}
\label{lem:factorial-upper-bound-primorial}
For all sufficiently large $n$, we have $n!\leq (n\ln n)\#$.
\end{Lem}

\begin{Proof}
Recall that for a positive integer $m$ we write write $m\#$ for the product of 
all 
prime numbers at most $m$. For $m\geq 563$, we have 
(cf.~Rosser and Schoenfeld~\cite{RosserS1962})
\[
\ln m\# > m\biggl(1-\frac{1}{2\ln m}\biggr)\,.
\]
For the factorial function, for $n\geq 1$, 
we have (cf.~Robbins~\cite{Robbins1955})
\[
n! = \sqrt{2\pi n}\biggl(\frac{n}{e}\biggr)^ne^{\alpha_n}\quad
\text{with}\quad\frac{1}{12n+1}<\alpha_n<\frac{1}{12n}\,,
\]
which gives us the comparatively crude upper bound, for $n\geq 1$, 
\[
\ln n! < \biggl(n+\frac{1}{2}\biggr)\ln n-n+1\,.
\]
We want $\ln n!<\ln m\#$. Accordingly, it suffices to have $m\geq 563$ and
\[
\biggl(n+\frac{1}{2}\biggr)\ln n-n+1<m\biggl(1-\frac{1}{2\ln m}\biggr)\,.
\]
It is immediate that $m\geq n\ln n$ suffices for $m\geq 563$, which completes the proof.
\end{Proof}

Thus, it suffices to work with all primes $p$ with $p\leq n\ln n$ in what follows.

\subsection{A Reduction from Zero-One Permanent to \#\textsc{InnerProduct}}

This section starts our work towards Theorem~\ref{thm:perm-to-exactip}
without yet parameterizing the reduction in detail.
Let a prime $2\leq p\leq n\ln n$ be given. 
We seek to compute $\per M$ modulo $p$.
Fix a primitive root $g\in\{1,2,\ldots,p-1\}$ modulo $p$. For 
an integer $a$ with $a\not\equiv 0\pmod p$, let us write
$\dlog_{p,g} a$ for the {\em discrete logarithm} of $a$ relative 
to $g$ modulo $p$. 
That is, $\dlog_{p,g} a$ is the unique integer in $\{0,1,\ldots,p-2\}$ that
satisfies $g^{\dlog_{p,g} a}\equiv a\pmod p$.
Working modulo $p$ and collecting the outer sum in \eqref{eq:ryser-recalled}
by the sign $\sigma\in\{-1,1\}$ and the {\em nonzero} products by their
discrete logarithm, we have
\[
\per M\equiv (-1)^n\sum_{e=0}^{p-2}\,
g^e\bigl(w_{1}^{(e)}-w_{-1}^{(e)}\bigr)
\pmod p\,,
\]
where
\[
w_{\sigma}^{(e)}
=\bigg|\biggl\{S\subseteq[n]\,:\,
(-1)^{|S|}=\sigma\ \ \text{and}\ 
\dlog_{p,g}\prod_{i\in [n]}\sum_{j\in S} m_{ij}\equiv e\!\!\pmod{p-1}
\biggr\}\bigg|
\]
for $\sigma\in\{-1,1\}$ and $e\in\{0,1,\ldots,p-2\}$. 
Thus, to compute $\per M$ modulo $p$ it suffices to compute the coefficients
$w_{\sigma}^{(e)}$. 

Towards this end, suppose that $n\geq 4$ is even and let 
\[
L=\{1,2,\ldots,n/2\}\qquad\text{and}\qquad
R=\{n/2,n/2+1,\ldots,n\}\,.
\]
For $\sigma_L,\sigma_R\in\{1,-1\}$, let
\[
w_{\sigma_L,\sigma_R}^{(e)}
=\bigg|\biggl\{\!S\subseteq[n]:
(-1)^{|S\cap L|}=\sigma_L\,,\,
(-1)^{|S\cap R|}=\sigma_R\ \ \text{and}\,
\dlog_{p,g}\prod_{i\in [n]}\sum_{j\in S} m_{ij}\equiv e\!\!\!\!\pmod{p-1}
\!\biggr\}\bigg|
\]
Clearly 
$w_{\sigma}^{(e)}=\sum_{\substack{\sigma_L,\sigma_R\in\{-1,1\}\\\sigma_L\sigma_R=\sigma}}w_{\sigma_L,\sigma_R}^{(e)}$,
so it suffices to focus on computing $w_{\sigma_L,\sigma_R}^{(e)}$ in what 
follows.
Define the set families
\[
\mathcal{L}_{\sigma_L}=\bigl\{A\subseteq L:(-1)^{|A|}=\sigma_L\bigr\}
\qquad
\text{and}
\qquad
\mathcal{R}_{\sigma_R}=\bigl\{B\subseteq R:(-1)^{|B|}=\sigma_R\bigr\}
\]
with $|\mathcal{L}_{\sigma_L}|=|\mathcal{R}_{\sigma_R}|=2^{n/2-1}$. 
Next we will define two families of length-$d$ zero-one vectors whose pair 
counts by inner product will enable us to recover the coefficients 
$w_{\sigma_L,\sigma_R}^{(e)}$. The structure of the vectors will be slightly
elaborate, so let us first define an index set $D$ for indexing the $|D|=d$ 
dimensions.
Let
\[
\begin{split}
D=\bigl\{&(i,\ell,r,k)\in [n]\times\{0,1,\ldots,p-1\}\times\{0,1,\ldots,p-1\}\times [np]:\\
&\qquad\ell+r\not\equiv 0\!\!\!\pmod p\ \text{implies} \ k\leq\dlog_{p,g}\bigl(\ell+r\bigr)\bigr\}\,.
\end{split}
\]
We have 
\[
d=n^2p^2+np(p-1)(p-2)/2<n^4(\ln n)^3\,. 
\]

For $A\in\mathcal{L}_{\sigma_L}$ and $B\in\mathcal{R}_{\sigma_R}$, 
define the vectors $\lambda(A)\in\{0,1\}^D$ and $\rho(B)\in\{0,1\}^D$ 
for all $(i,\ell,r,k)\in D$ by the rules
\begin{equation}
\label{eq:lambda-rho}
\lambda(A)_{i\ell rk}=
\begin{cases}
1 & \!\!\!\text{if $\ell\equiv\sum_{j\in A}m_{ij}\!\!\!\pmod p$;}\\
0 & \!\!\!\text{otherwise;}
\end{cases}\ \ \text{and}\ \ 
\rho(B)_{i\ell rk}=
\begin{cases}
1 & \!\!\!\text{if $r\equiv\sum_{j\in B}m_{ij}\!\!\!\pmod p$;}\\
0 & \!\!\!\text{otherwise.}
\end{cases}
\end{equation}

To study the inner product $\bra\lambda(A),\rho(B)\ket$ it will be convenient
to work with Iverson's bracket notation. Namely, for a logical proposition $P$, 
let
\[
\iv{P}=\begin{cases}
1 & \text{if $P$ is true};\\
0 & \text{if $P$ is false}.
\end{cases}
\]
Over the integers, from \eqref{eq:lambda-rho} we now have
\begin{equation}
\label{eq:lambda-rho-ip}
\begin{split}
\bra\lambda(A),\rho(B)\ket
&=
\sum_{(i,\ell,r,k)\in D}
\lambda(A)_{i\ell rk}
\rho(B)_{i\ell rk}\\
&=
\sum_{(i,\ell,r,k)\in D}
\iv{\ell\equiv\sum_{j\in A}m_{ij}\!\!\!\pmod p}
\iv{r\equiv\sum_{j\in B}m_{ij}\!\!\!\pmod p}\\
&=
\sum_{i\in[n]}
\hspace{-18mm}
\sum_{\hspace{17mm}\tiny\begin{array}{l}\ell=0\\\ell+r\not\equiv 0\!\!\!\!\pmod p\end{array}}^{p-1}\hspace{-17mm}
\sum_{r=0}^{p-1}\,\,\,
\iv{\ell\equiv\sum_{j\in A}m_{ij}\!\!\!\pmod p}
\iv{r\equiv\sum_{j\in B}m_{ij}\!\!\!\pmod p}
\dlog_{p,g}\bigl(\ell+r\bigr)
\\
&\qquad+\sum_{i\in[n]}
\sum_{\ell=0}^{p-1}
\iv{\ell\equiv\sum_{j\in A}m_{ij}\!\!\!\pmod p}
\iv{p-\ell\equiv\sum_{j\in B}m_{ij}\!\!\!\pmod p}
np\\
&=\begin{cases}
\sum_{i\in[n]}\dlog_{p,g}\sum_{j\in A\cup B}m_{ij}
& \text{if $\prod_{i\in [n]}\sum_{j\in A\cup B}m_{ij}\not\equiv 0\!\!\!\pmod p$;}\\
\geq np 
& \text{if $\prod_{i\in [n]}\sum_{j\in A\cup B}m_{ij}\equiv 0\!\!\!\pmod p$.}
\end{cases}
\end{split}
\end{equation}
In particular, letting 
\[
f_{\sigma_L,\sigma_R,t}=
\big|\bigl\{(A,B)\in\mathcal{L}_{\sigma_L}\times\mathcal{R}_{\sigma_R}:
  \bra\lambda(A),\rho(B)\ket=t\bigr\}\big|\,,
\]
it follows immediately from \eqref{eq:lambda-rho-ip} that we have 
$w_{\sigma_1,\sigma_2}^{(e)}
 =\sum_{t=0,\ t\equiv e\!\pmod{p-1}}^{n(p-2)}f_{\sigma_L,\sigma_R,t}$,
which enables us to recover $\per M$ from the counts of pairs
in $\mathcal{L}_{\sigma_L}\times\mathcal{R}_{\sigma_R}$ by inner product.

\subsection{Completing the Proof of Theorem~\ref{thm:perm-to-exactip}}

Suppose we have an algorithm for \#\textsc{InnerProduct} that
runs in $N^{2-\Omega(1/\log c)}$ time when given an input of $N$
vectors from $\{0,1\}^{c\log N}$. Take $N=2^{n/2-1}$ and observe that
$\log N=n/2-1$. The reduction from previous section has $d\leq n^4(\ln n)^3$
and thus we can take $c=(n\ln n)^3$ and thus solve $n\times n$ 
zero-one permanent in time $N^{2-\Omega(1/\log c)}=2^{n-\Omega(n/\log n)}$. 
This completes the proof of Theorem~\ref{thm:perm-to-exactip}.

\section{A Lower Bound for \#\textsc{OV} via Zero-One Permanents}

\label{sect:perm-cont}

This section continues our work towards relations to zero-one permanents
started in Sect.~\ref{sect:perm}; in particular, we 
prove Theorem~\ref{thm:perm-to-ov}.

\subsection{A Reduction from Zero-One Permanent to \#OV}

This section starts our work towards Theorem~\ref{thm:perm-to-ov}
without yet parameterizing the reduction in detail.
As in Sect.~\ref{sect:perm}, it suffices to describe how to compute $\per M$
modulo a given prime $p$ with $2\leq p\leq n\ln n$. 

Let $g$ be a positive integer parameter, which we assume divides $n$. 
For $h\in[g]$, let
\[
V_h=\{i\in [n]:(h-1)n/g+1\leq i\leq hn/g\}
\]
be a partition of the rows of $M$ into $g$ groups, each of size $n/g$.
Again from Ryser's formula, we observe that
\[
\per M=(-1)^n\sum_{S\subseteq[n]}(-1)^{|S|}\prod_{h\in[g]}\prod_{i\in V_h}\sum_{j\in S}m_{ij}\,.
\]
Grouping by sign $\sigma\in\{-1,1\}$ and per-group residues 
$r\in\{0,1,\ldots,p-1\}^g$ modulo $p$, we thus have
\begin{equation}
\label{eq:g-group-ryser}
\per M\equiv(-1)^n\!\!\!\sum_{r\in\{0,1,\ldots,p-1\}^g}\!\!\!(t_{1,r}-t_{-1,r})\prod_{h=1}^g r_h\pmod p\,,
\end{equation}
where
\[
t_{\sigma,r}=
\big|\bigl\{S\subseteq[n]:
(-1)^{|S|}=\sigma\ \text{and}\ 
\prod_{i\in V_h}\sum_{j\in S}m_{ij}
\equiv
r_h
\!\!\!
\pmod{p}\ 
\text{for each $h\in[g]$}
\bigl\}\big|\,.
\]
Observe that given all the counts $t_{\sigma,r}$, it takes $O(p^{g}g)$ 
operations modulo $p$ to compute the permanent modulo $p$
via \eqref{eq:g-group-ryser}, which is less than $2^nn$ when $g<n/\log p$.
We continue to describe how to get the counts $t_{\sigma,r}$
via orthogonal-vector counting.

Assuming that $n\geq 4$ is even, introduce again the split 
\[
L=\{1,2,\ldots,n/2\}\qquad\text{and}\qquad
R=\{n/2,n/2+1,\ldots,n\}\,.
\]
Let the residue vector $r\in\{0,1,\ldots,p-1\}^g$ be fixed.
For $\sigma_L,\sigma_R\in\{1,-1\}$, let
\[
\begin{split}
t_{\sigma_L,\sigma_R,r}=
\big|\bigl\{S\subseteq[n]:\ &
(-1)^{|S\cap L|}=\sigma_L\ ,\ 
(-1)^{|S\cap R|}=\sigma_R\,,\\&\quad\text{and}\ 
\prod_{i\in V_h}\sum_{j\in S}m_{ij}
\equiv
r_h
\!\!\!
\pmod{p}\ 
\text{for each $h\in[g]$}
\bigl\}\big|\,.
\end{split}
\]
Clearly 
$t_{\sigma,r}=\sum_{\substack{\sigma_L,\sigma_R\in\{-1,1\}\\\sigma_L\sigma_R=\sigma}}t_{\sigma_L,\sigma_R,r}$,
so it suffices to focus on computing $t_{\sigma_L,\sigma_R,r}$ in what 
follows. We again work with the set families
\[
\mathcal{L}_{\sigma_L}=\bigl\{A\subseteq L:(-1)^{|A|}=\sigma_L\bigr\}
\qquad
\text{and}
\qquad
\mathcal{R}_{\sigma_R}=\bigl\{B\subseteq R:(-1)^{|B|}=\sigma_R\bigr\}\,.
\]
Let
\[
D=[g]\times\{0,1,\ldots,p-1\}^{n/g}\,.
\]
We have 
\[
d=|D|=gp^{n/g}\,.
\]

For $A\in\mathcal{L}_{\sigma_L}$ and $B\in\mathcal{R}_{\sigma_R}$, 
define the vectors $\lambda(A)\in\{0,1\}^D$ and $\rho(B)\in\{0,1\}^D$ 
for all $(h,u)\in D$ by the rules
\begin{equation}
\label{eq:lambda-rho-ov}
\begin{split}
\lambda(A)_{hu}&=
\begin{cases}
1 & \!\!\!\text{if we have $\sum_{j\in A}m_{ij}\equiv u_{i-(h-1)n/g}\!\!\!\pmod p$ for all 
                   $i\in V_h$;}\\
0 & \!\!\!\text{otherwise;}
\end{cases}\\[3mm]
&\hspace{-12mm}\text{and}\\[3mm]
\rho(B)_{hu}&=
\begin{cases}
0 & \!\!\!\text{if $\prod_{i\in V_h}\bigl(u_{i-(h-1)n/g}+\sum_{j\in B}m_{ij}\bigr)\equiv r_h\!\!\!\pmod p$;}\\
1 & \!\!\!\text{otherwise.}
\end{cases}
\end{split}
\end{equation}
Over the integers, from \eqref{eq:lambda-rho-ov} we now have
\begin{equation}
\label{eq:lambda-rho-ip-ov}
\begin{split}
\bra\lambda(A),\rho(B)\ket
&=\sum_{(h,u)\in D}
\lambda(A)_{hu}
\rho(B)_{hu}\\
&=\sum_{h\in[g]}
  \sum_{u\in\{0,1,\ldots,p-1\}^{n/g}}
  \prod_{i\in V_h}
  \bbiv{\sum_{j\in A}m_{ij}\equiv u_{i-(h-1)n/g}\!\!\!\pmod p}\\
&\hspace*{37mm}
  \bbiv{\prod_{i\in V_h}\bigl(u_{i-(h-1)n/g}+\sum_{j\in B}m_{ij}\bigr)\not\equiv r_h\!\!\!\pmod p}\\
&=\sum_{h\in[g]}
  \bbiv{\prod_{i\in V_h}\biggl(\sum_{j\in A}m_{ij}+\sum_{j\in B}m_{ij}\biggr)\not\equiv r_h\!\!\!\pmod p}\\
&=\begin{cases}
0      & 
\text{if we have $\prod_{i\in V_h}\sum_{j\in A\cup B}m_{ij}\equiv r_h\!\!\! \pmod{p}$ for each $h\in[g]$;}\\
\geq 1 & \text{otherwise.}
\end{cases}
\end{split}
\end{equation}
In particular, we have
\[
t_{\sigma_L,\sigma_R,r}=
\big|\bigl\{(A,B)\in\mathcal{L}_{\sigma_L}\times\mathcal{R}_{\sigma_R}:
  \bra\lambda(A),\rho(B)\ket=0\bigr\}\big|\,,
\]
which enables us to recover $\per M$ from the counts of orthogonal pairs
in $\mathcal{L}_{\sigma_L}\times\mathcal{R}_{\sigma_R}$.

\subsection{Completing the Proof of Theorem~\ref{thm:perm-to-ov}}

Suppose now that we have an algorithm for \#\textsc{OV} that
runs in $N^{2-\Omega(1/\log^{1-\epsilon} c)}$ time for some $0<\epsilon<1$
when given an input of $N$
vectors from $\{0,1\}^{c\log N}$. Take $N=2^{n/2-1}$ and observe that
$\log N=n/2-1$. 

Let $K>1$ be a constant that will depend on $\epsilon$ and the constant
hidden by the $\Omega(\cdot)$ notation in the running time of the
\#\textsc{OV} algorithm. Take
\[
g=\lfloor K^{-1/\epsilon}n(\log p)^{1-2/\epsilon}\rfloor
\]
and recall that the prime $p$ is in the range $2\leq p\leq n\ln n$.
To compute the parameters $t_{\sigma,r}$ using the reduction in the previous
section, for each prime $p$ we need $4p^g$ invocations of the \#\textsc{OV}
algorithm on an input of $N$ vectors of dimension $d=gp^{n/g}$.
Thus, for all large enough $n$, 
since $\frac{1}{2}K^{-1/\epsilon}n(\log p)^{1-2/\epsilon}\leq g$,
we have 
\[
d=gp^{n/g}
\leq n 2^{2K^{1/\epsilon}(\log p)^{2/\epsilon}}\,.
\]
Since clearly $d=c\log N=c(n/2-1)$ and $2/\epsilon>2$, 
for all large enougn $n$, we have 
\[
\begin{split}
\log c&\leq 1+2K^{1/\epsilon}(\log p)^{2/\epsilon}\\
      &\leq 3K^{1/\epsilon}(\log p)^{2/\epsilon}\,,
\end{split}
\]
where the last inequality depends on choosing a large enough $K$ 
so that the inequality is true for $p=2$. 
Thus,
\[
\begin{split}
-(\log c)^{\epsilon-1}
&\leq -{3^{\epsilon-1}K^{1-1/\epsilon}(\log p)^{2-2/\epsilon}}\,.
\end{split}
\]
One invocation of the \#\textsc{OV} algorithm thus runs in 
\[
N^{2-\Omega(\log^{\epsilon-1} c)}=2^{n-\Omega(n{3^{\epsilon-1}K^{1-1/\epsilon}(\log p)^{2-2/\epsilon}})} 
\]
time. For each prime $2\leq p\leq n\ln n$, we need
\[
4p^g\leq 2^{2+K^{-1/\epsilon}n(\log p)^{2-2/\epsilon}}
\]
invocations of the \#\textsc{OV} algorithm. Thus, the running time of 
all the invocations for the prime $p$ is bounded by
\[
\begin{split}
4p^gN^{2-\Omega(\log^{\epsilon-1} c)}\leq
2^{n-\Omega(n{3^{\epsilon-1}K^{1-1/\epsilon}(\log p)^{2-2/\epsilon}})
   +2+K^{-1/\epsilon}n(\log p)^{2-2/\epsilon}}\,.
\end{split}
\]
By choosing a large enough $K$ to dominate the constant hidden by 
the $\Omega(\cdot)$ notation in the running time of 
the \#\textsc{OV} algorithm, we thus have, for all large enough $n$,
\[
\begin{split}
4p^gN^{2-\Omega(\log^{\epsilon-1} c)}
&\leq 
2^{n-\Omega(n{3^{\epsilon-1}K^{-1/\epsilon}(\log p)^{2-2/\epsilon}})}\\
&\leq 
2^{n-\Omega(n{3^{\epsilon-1}K^{-1/\epsilon}(\log n+\log\ln n)^{2-2/\epsilon}})}\\
&\leq
2^{n-\Omega(n(\log n)^{2-2/\epsilon})}\,.
\end{split}
\]
Since there are at most $n\ln n$ primes $p$ to consider, the total running
time to compute $\per M$ is bounded by $2^{n-\Omega(n/\log^{2/\epsilon-2}n)}$.
This completes the proof of Theorem~\ref{thm:perm-to-ov}.

\section{Further Applications}

\subsection{Counting Satisfying Assignments to a \textsc{Sym}$\circ$\textsc{And} circuit via \#\textsc{InnerProduct}}

\label{sect:sym-and}

We describe how to embed a \textsc{Sym}$\circ$\textsc{And} circuit, i.e., a circuit of $s$ \textsc{And} gates working on $n$ Boolean inputs, connected by a top gate that is an arbitrary symmetric gate, in a \#\textsc{InnerProduct} instance of size $N=2^{n/2}$ and $d=s$. Assuming $n$ even, we divide the $n$ inputs in two equal halves $L$ and $R$. We let $\mathcal A$ have one vector $u$ for each assignment to the inputs in $L$, with one coordinate in $u$ for each \textsc{And} gate, representing the truth value of that
gate restricted to the inputs in $L$. Likewise, we let $\mathcal B$ have one vector $v$ for each assignment to the inputs in $R$, with each coordinate set to the truth value of the represented gate restricted to the inputs in $R$. It is readily verified that $\langle u,v\rangle$ counts the number of \textsc{And} gates that are satisfied by the assignment represented by $(u,v)$.
Hence, knowing the number of assignments that satisfy exactly $t$ of the \textsc{And} gates,  for $t=0,1,\ldots,s$, which is what the solution to the \#\textsc{InnerProduct} gives us, we can count the total number of assignments that also satisfies the top symmetric gate.

Variations where the circuit instead is a \textsc{Sym}$\circ$\textsc{Or} or a \textsc{Sym}$\circ$\textsc{Xor}, are also possible.  

\subsection{Computing the Weight Enumerator Polynomial via \#\textsc{InnerProduct}}

\label{sect:weight-enum}

A binary linear code of length $n$ and rank $k$ is a linear subspace $C$ with dimension $k$ of the vector space 
$\mathbb{F}_2^n$. 
The \emph{weight enumerator polynomial} is
\[
W(C;x,y)=\sum_{w=0}^n A_wx^wy^{n-w}\,,
\]
where
\[
A_w=|\{c\in C:\langle c,c\rangle=w\}|\,,
\]
for $w=0,1,\ldots,n$ is the {\em weight distribution}; that is, 
$A_w$ equals the number of codewords of $C$ having exactly $w$ ones.

We will reduce the computation of the weight distribution, and hence the weight enumerator polynomial, to $(k/2+1)^2$ instances of \#\textsc{InnerProduct} with $N\leq 2^{k/2}$ and $d=2(n-k)$ when $k$ is even.

Let the $k\times n$ matrix $G$ be the generating matrix of the code; that is, the codewords of $C$ are exactly the row-span of $G$. We can assume without loss of generality that the generator matrix has the standard form $G=[I_k|P]$, where $I_k$ is the $k\times k$ identity matrix. 
For each $s_A=0,1,\cdots,k/2$ and $s_B=0,1,\ldots,k/2$, we make one instance 
of \#\textsc{InnerProduct}.

We let the set $\mathcal A$ have one vector $u$ for each code $c$ obtained as the linear combination of exactly $s_A$ of the first $k/2$ rows. Each of the $n-k$ last coordinates in the code word $c$ is described by a block of two coordinates in $u$. If $c_i=0$ we encode this as $01$ in $u$, and if $c_i=1$ we encode this as $10$ in $u$. We concatenate all $n-k$ encoded blocks to obtain $u$. Likewise, we let the set $\mathcal B$ have one vector $v$ for each code $c$ obtained as a linear combination of $s_B$ of the last $k/2$ rows. Again,
each of the $n-k$ last coordinates in the code word $c$
is described by a block of two coordinates in $v$, but the encoding is opposite the one for $\mathcal A$: If $c_i=0$ we encode this as $10$ in $v$, and if $c_i=1$ we encode this as $01$ in $v$. We again concatenate all $n-k$ encoded blocks to obtain $v$.
With this design, it is readily verified that for $(u,v)\in \mathcal A\times \mathcal B$, the inner product $\langle u,v\rangle$ is equal to the number of ones in the last $n-k$ coordinates in the code word 
obtained as the sum of the code word represented by $u$ and the code word represented by $v$. Also, by design the number of ones in the first $k$ coordinates equals $s_A+s_B$. Hence, by summing over all pairs that have the same inner product $t$, and aggregating over all $s_A$ and $s_B$, we can compute the weight distribution.


\section*{Acknowledgment}

We thank Virginia Vassilevska Williams and Ryan Williams for many useful discussions.


\bibliographystyle{abbrv}
\bibliography{paper.bib}


\end{document}